\documentclass[12pt]{article}

\usepackage{latexsym}
\usepackage{amssymb}

\usepackage[francais,german, english]{babel}
\def\nn{\nonumber}
\def\be{\begin{equation}} 
\def\ee{\end{equation}} 
\def\ba{\begin{eqnarray}} 
\def\ea{\end{eqnarray}} 
\def\ka{\kappa^2}
\def\half{\frac{1}{2}}

\def\th{\tilde{h}}
\def\tk{\tilde{k}}
\def\tF{\tilde{F}}
\def\tp{\tilde{\phi}}
\def\deriv{\mathrm{d}}
\def \pa {\partial}

\def \la {\lambda}
\def \La {\Lambda}
\def \Da {\Delta}

\def \a {\alpha}

\def \da {\delta}
\def \ep {\epsilon}

\def \om {\omega}

\begin{document}

\begin{center}
\vspace{1cm}
{ \bf  \Large Wave function of the radion with  BPS branes} 

\vspace{1cm}
\large{ G.\ Cynolter\footnote{ \noindent 
Leave on absence from HAS Research 
Group for Theoretical Physics, Budapest} 
\footnote{ E-mail: cynoter@itp.phys.ethz.ch}
}

\vspace{0.35cm}
\normalsize
{\it Institute of Theoretical Physics, 
ETH \\ CH-8093 Z\"urich, Switzerland}

\end{center}

\abstract{We investigate coupled gravity and a bulk scalar field
 on a slice  of AdS$_5$ bulk with special  BPS branes at the two ends.
With the special scalar potentials on the branes the scalar field
does not stabilize the size of the orbifold.
With a careful treatment of the general coordinate invariance
the complete tensor-scalar spectrum is  presented.
There are two massless zero modes in
the scalar sector, the radion and a dilatonic zero mode.
The scalar KK modes acquire masses at the order of the 
warped mass scale. The four dimensional effective
action is of tensor-scalar type.
}

\section{Introduction}

Recently there has been a growing interest in extra dimensional models
where the standard model fields localized on a brane. 
Randall and Sundrum proposed a scenario (RS1) with one extra dimension,
two branes of  opposite tension at the orbifold points of an AdS bulk
\cite{Randall:1999vf}.
This setup was motivated by the recent progress in string/M-theory
and can be realized in string theory 
\cite{Verlinde:1999fy}, \cite{Giddings:2001yu}.
Poincare invariant solution requires a finetuning of the bulk
cosmological constant $\La= -\frac{6k^2}{\ka}$ and the brane tensions
$\la_{Planck}=-\la_{\rm{TeV}}=\frac{6k}{\ka}$.
The exponentially warped metric can generate
the hierarchy between the Planck and the electroweak scale with
moderate value parameters $k r_c \sim {\cal O} (40)$.
However, there is an exact solution for any $r_c$ distance of
the two branes. A change in $r_c$ is described by a four dimensional
massless scalar field, the radion \cite{cgr}.
It couples to matter like a Brans-Dicke scalar
and must have a large Brans-Dicke parameter or a
mass to recover standard 4-dimensional gravity \cite{Csaki:1999mp},
\cite{Garriga:2000yh}.

Goldberger and Wise proposed a mechanism with a massive 
bulk scalar field  to stabilize the radion \cite{Goldberger:1999uk} .
They added special steep $\la_i(\Phi)$ brane potentials.
The competition between the kinetic and potential terms 
generates  a minimum in the effective potential and determines
$r_c$. Approaches based on naive ansatz without including the 
backreaction found that the radion acquires an ${\cal O}$(TeV) mass 
and the couplings  to standard model fields are $\sim1/$TeV
\cite{Csaki:1999mp}.
Effective field theory calculation gave the same radion mass
in \cite{Goldberger:1999un}.
Tanaka, Montes \cite{Tanaka:2000er}  and later independently
Cs\'aki et al. included the backreaction and derived a single
equations for the scalar degrees of freedom based on Lorentz covariance
of the linear perturbations \cite{Csaki}.
The equation together with the boundary conditions at the branes
defines a hermitic operator only in special cases, like the stiff potential
limit, when $\frac{\pa^2 \la_i}{\pa \Phi^2} \rightarrow \infty$.
In this limit the physical results agree with the naive calculations.

Another mathematically well defined case is when
the brane potential is equivalent to the superpotential that
generates the bulk scalar potential (\ref{bpscond}), 
we call these BPS branes.
This type of models  with  exponential potentials 
can possibly arise from the bosonic sector of  dimensionally 
reduced supergravity theory \cite{Cvetic:2000pn}.
In this letter we give the complete spectrum of
linear perturbations around an exact background solution
of two flat BPS branes in 5 dimensions.
This scenario was analyzed in \cite{Brax:2002kv} 
using two different Gaussian normal coordinate patches
with respect to the two branes following 
the method of \cite{cgr}.
They discussed only one scalar perturbation 
the massless radion.
Solving the equations with two flat branes we find
still one more massless dilatonic scalar and the KK tower
for the scalar field.
This model does not solve the moduli problem but
the complete perturbative spectrum can be understood, the limit
to the Randall-Sundrum case is smooth and we hope that
it gives help to solve the general case with detuned brane potentials.
The low energy effective theory is of scalar-tensor nature
and generally it cannot be accepted by phenomenology and cosmology.

The paper is organized as follows.
In section 2. we present the setup and discuss the background
solution using the technique of \cite{DeWolfe:2000cp}.
In section 3. we consider  the simple limit of the 
Randall Sundrum scenario with an additional bulk scalar field.
In section 4. we fix the gauge, solve the linearized equations
and present the complete spectrum and we conclude
in the last section.

\section{The model}

In a 5 dimensional  ( $ x^a=(x^{\mu},y=x^5) $ )
$M_4 \times S_1/Z_2$ manifold, there are scalar field, gravity and
two flat branes at  the orbifold fixed points $y=0, r_c$.
Latin indices run from 1..5 and greek indices 1..4 and
prime denotes the derivation with respect to the proper coordinate $y$.
4-dimensional greek indices raised and lowered by $\eta_{\mu \nu}$.
The action we consider is 
\begin{equation}
S= \int d^5x \sqrt{g} 
\left (-M^3R+\half g^{ab} \nabla_a \Phi \nabla_b \Phi -V(\Phi)\right ) 
    - \sum_{i=P,T} \int d^4x \sqrt{g_{i}} \lambda_i(\Phi) ,
\label{action}
\end{equation}
$g_{i}$ 
is the induced metric on the brane, $ \ka=1/(2M^3)$ 
is the 5D cosmological constant,
the integration understood on the orbifold.

We want to investigate perturbations around a background solution.
The ansatz for solving the Einstein scalar coupled system
that respects Lorentz invariance is
\ba
ds^2 &=& a^2(y) \eta_{\mu\nu} dx^{\mu} dx^{\nu}   - dy^2,  \\
\Phi(x,y) &=& \Phi_0(y).
\ea
The warp factor $a(y)$ and $\Phi_0(y)$ are continuous even functions. 
The first derivatives may have
jumps, the second derivatives delta singularities at the branes.
We use the following form of the Einstein equations
\be
 R_{ab}=\ka \tilde{T}_{ab}=\ka \left 
({T}_{ab}-\frac{1}{3}g_{ab}g^{cd}T_{cd}   \right ).  \nn
\ee
The gravitational and scalar  equations are the following.
\begin{eqnarray}
4 \left (\frac{a'}{a} \right )^2+  \left ( \frac{a'}{a}  \right)' & =& - \frac{2 \ka}{3}
V(\Phi_0)-\frac{\ka}{3} \sum_i \lambda_i \delta(y-y_i)  \label{einstein1} \\
\left ( \frac{a'}{a} \right )^2 & = & - \frac{\ka}{6}V(\Phi_0)+
\frac{\ka}{12}\Phi_0'^2  \label{einstein2} \\
\Phi_0'' & = & -4 \frac{a'}{a} \Phi_0' + \frac{\partial V(\Phi_0)}{\partial \Phi}
+\sum_i \frac{ \partial \lambda_i}{\partial \Phi}  \delta(y-y_i)  
\end{eqnarray}
The boundary conditions are
\be
\left [ \frac{a'}{a}  \right ]_{y_i}=-\frac{\ka}{3}\lambda_i, 
\quad 
\left [\phantom{\frac{a'}{a}}
{\Phi_0'} \right ]_{yi}=\frac{\partial \lambda_i}{\partial \Phi}
. \label{bc1}
\ee
With $Z_2$ symmetry and working in the $(0,r_c)$ interval
the jump is twice the value of the derivative calculated from the
inside of the interval and there is sign difference at the two branes.

An analytic solution of these equation is generated by the 
technique proposed by DeWolfe et al.\cite{DeWolfe:2000cp}.
With one $W(\Phi)$ superpotential a flat solution is generated via
\ba
\Phi_0' &=& \half \frac{\partial W}{\partial \Phi}, \\
\frac{a'}{a} &=& - \frac{\ka}{6} W(\Phi_0), \\
V(\Phi_0) &=& \frac{1}{8} \left (\frac{ \partial W(\Phi_0)}{\partial \Phi}
\right )^2 - \frac{\ka}{6} W(\Phi_0)^2 .
\ea
These solve the equations of motions if the boundary conditions for $W(\Phi)$
are fulfilled (+ for $y=0$ and - for $y=r_c$, $\Phi_i=\Phi_0(y_i)$).
\ba
\lambda_i(\Phi_i) &=& \pm W(\Phi_i) \cr
\frac{\deriv \lambda_i}{\deriv \Phi} (\Phi_i) &=& 
\pm \frac{\deriv W}{\deriv \Phi} (\Phi_i)  \label{tune}
\ea
Exactly one fine tuning is needed to find a flat background solution.
There is no need for supersymmetry to find a solution
though the bulk plus brane action completely vanishes substituting
the solution.
This technique reduces  the coupled system of differential equation
to decoupled ordinary differential equations and can be solved
(generally numerically) for arbitrary $V(\Phi)$ and $\la_1(\Phi)$
tuning one additive constant in $\la_2(\Phi)$.
Explicit solution can be constructed defining first $W(\Phi)$,
like $W(\Phi)= c-u \Phi^2$ or $W(\Phi)=e^{\alpha \Phi}$.

We will investigate perturbations in  models where the fine tuning
of the scalar brane- and superpotential  (\ref{tune}) 
is valid at all order for arbitrary $\Phi$.
\be
\lambda_i(\Phi)= \pm W(\Phi)
\label{bpscond}
\ee
We call this type of branes BPS branes. 
These branes allow solutions in the bosonic sector 
respecting 5 dimensional bulk+brane supersymmetry, the BPS condition
preserves half of the supercharges corresponding to 
unbroken N=1 supersymmetry in 4 dimensions
\cite{Bergshoeff:2000zn}.
This type of models with  exponential potentials can possibly arise from
the bosonic sector of  dimensionally reduced supergravity theory
\cite{Cvetic:2000pn}. The localization of scalar modes were
discussed with one brane in \cite{Bozza:2001xt} based on 
$O(3)$ symmetry in the generalized longitudinal gauge of
\cite{vandeBruck:2000ju}.

There is always a flat background solution for any value of $r_c$. 
The change of $r_c$ is described by a massless modulus field,
the radion in the low energy effective theory.
There is an additional dilatational invariance
of the action with BPS branes leading to a second zero mode.
The value of the scalar field on the Planck brane
is an integration constant, a free parameter of the action.
Choosing a definite background solution this symmetry is broken
spontaneously and the corresponding massless Goldstone boson is 
expected in the scalar spectrum.
In the limit of the Randall-Sundrum scenario this invariance
is obvious and is presented in the next section.

To obtain an analytical solution for the KK modes the following
particular potential will be used
\be
W(\Phi)=\pm\la_i(\Phi)= \frac{6 k}{\ka} e^{\a \Phi}.
\label{dilatonpot}
\ee
The solution contains two integration constants $c_0, C_P$, usually both
chosen to $1$
\ba
a(y)&=& c_0(C_P-\frac{3 \a^2}{\ka}k y )^{\frac{\ka}{3 \a^2}} ,\\
e^{-\a \Phi_0(y)}&=& (C_P-\frac{3 \a^2}{\ka}k y ).
\ea
In the limit $\a \rightarrow 0$ 
we get back the AdS $e^{-k y}$ profile of the 
Randall-Sundrum scenario and the scalar field decouples from gravity.
We deal with solution with $0< \a^2< \frac{\ka}{3kr_c}$ 
to avoid naked singularities in
the bulk. This solution generates a hierarchy of
$a(r_c)= (1-\frac{3 \a^2}{\ka}k r_c )^{\frac{\ka}{3 \a^2}}$.
The dilatational invariance  corresponds to
the free choice of  $C_P=e^{-\a \Phi_0(0)}$ 
the value of the scalar field on the Planck brane. 
It's change can be compensated by a change in $k,c_0$ to have 
the same warp factor $a(y)$ and superpotential.
For particular solutions we use $c_0=C_P=1$.
In the next section we discuss the special limit
of the Randall-Sundrum scenario.

\section{RS1  with bulk scalar field}

In the RS1 scenario specially tuned bulk and brane cosmological
constants provide a warped solution. In a more realistic model
a dynamical mechanism is needed to produce the brane and bulk constants.
A simple proposal is to add a bulk scalar field which has constant 
bulk and brane potentials.
This is the limit of the previously presented model with
\ba
&& W(\Phi)=\la(\Phi)=\frac{6 k}{\ka}= \hbox{constant}, \\
&& V(\Phi)= \La=-\frac{6 k^2}{\ka}.
\ea
The classical background solution 
\be
a(y)= e^{-k|y|} , ~ ~ \Phi_0(y)=\Phi_{RS}= \hbox{constant}
\label{RSscalar}
\ee 
has two invariances. One is related to the
free choice of the distance $r_c$ between the two branes,
the other is the freedom that we can shift the constant 
scalar field $\Phi_{RS}$ to $\Phi_{RS}+c_1$, it is also an 
exact solution with the same warp factor.

In the linearized theory the former provides the massless radion
\cite{cgr} the latter zero mode can be identified as follows.
The linearized gravitational ($h_{ab}$) and scalar ($\phi(x,y)$) 
perturbations around (\ref{RSscalar}) decouple,
it was also realized in \cite{Bozza:2001xt},\cite{sundrum}.
The gravitational perturbations are as in RS1, a massless
radion with the wavefunction
\be
F(x,y)= - \frac{a'}{a} \frac{1}{a^2}f(x)= k \frac{1}{a^2}f(x),
~ \Box f(x)=0,
\ee
 a massless spin-2 graviton localized on the Planck brane
and its KK tower.

\noindent 
The scalar field perturbations are determined from 
\be
\Box_5 \phi(x,y)=0,
\ee
with the boundary condition $\phi'(x,y_i)=0$.
This is the  free Klein-Gordon equations on an AdS$_5$ background,
with $\Box=\Box_4= \pa^\mu \pa_\mu$
\be
\Box_5= -\pa_z^2- 4 \frac{a'}{a} \pa_z + \frac{1}{a^2} \Box.
\ee
The scalar field spectrum contains a dilaton zero mode
\be
\phi(x,y)=f_1(x), ~~ \Box f_1(x)=0,
\label{RSdilaton}
\ee
reflecting the invariance of the action against the shift of
the constant value of the scalar field.
The background solution brakes this symmetry spontaneously
and this zero mode is the corresponding Goldstone boson.
This zero mode mixes with gravitational perturbations
if the linearized scalar equations are coupled to gravity.

We solve the equation for the  KK modes in Schr\"odinger basis:
change to conform flat background
by $dy= a(z) dz$ with $a(z)=1/(1+k|z|)$ and  rescale the field 
by $\phi(x,z)=e^{ipx} \Phi_0' a^{-\frac{3}{2}} \tp(z)$
also changing to mass eigenvectors with
$\Box \phi_i=-m_i^2 \phi_i$.
\be
\tp''-\left( \frac{9}{4} \left ( \frac{a'}{a} \right )^2+
\frac{3}{2}\left ( \frac{a'}{a} \right)'  \right ) \tp + m^2 \tp=0
\ee
The boundary condition is $(a^{-\frac{3}{2}} \tp(z))'=0$.
The potential term multiplying $\tp$ can be written as
\be
V(z)= \frac{\left (a^{\frac{3}{2}} \right )''}{a^{\frac{3}{2}}}=
\frac{(4-\frac{1}{4} )k^2}{(1+k |z|)^2}.
\ee
The massive solutions are determined by  Bessel functions
\be
\tp = \sqrt{1+k|z|} \left (  
A J_2\left(\frac{m}{k}(1+k|z|)\right)
+B Y_2\left (\frac{m}{k}(1+k|z|)\right )  \right ).
\label{RSphi}
\ee
The ratio $A/B$ determined from one boundary condition
the mass eigenvalues are determined at the other brane and
are of the order of the TeV scale.

The scalar spectrum then contains two zero modes and
a KK tower. Turning on smoothly the bulk and brane scalar
field potentials continuous deformation of the spectrum is 
expected. 
Quantum corrections are also expected to develop
a nontrivial bulk and brane potential for the scalar field.
The linearized  equations for the action  (\ref{action})
is presented in the next section.

\section{Linear perturbations}

In this section we give the equation of motion
for general linear perturbations keeping the branes 
at the fixed orbifold points and discuss the role of
gauge freedom.
A general perturbation is
\ba
&&\da g_{\mu \nu}=a^2 h_{\mu \nu}(x,y),~~ 
\da g_{5 \nu}=h_{5 \nu}(x,y)  \cr
&&\da g_{55} = h_{55}=-2G(x,y), \cr
&& \da \Phi=\phi(x,y).
\ea
Infinitesimal  general coordinate transformations
\be
x^5\rightarrow x^5+\xi^5, ~~~ x^\mu \rightarrow x^\mu+\xi^\mu,
\ee
must fulfill the orbifold symmetry conditions,
$\xi_\mu$ continuous and even, $\xi_5$ odd and vanishes
at the branes.
\ba
\xi_\mu(x,y+y_i)=\xi_\mu(x,y-y_i) \\
\xi_5(x,y+y_i)=-\xi_5(x,y-y_i)  \label{xi5orb}
\ea
$\xi_5$ gauge transformations moving the orbifold fixed points
are not allowed in this approach.
The  infinitesimal change of the background solution 
is understood as a change in the perturbations.
\ba
\delta h_{\mu \nu} &=& ( \xi^{\hbox{} }_{\mu, \nu}+
\xi^{\hbox{} }_{\nu, \mu} ) - 2\frac{a'}{a} \xi_5 \eta_{\mu \nu} \cr
\delta h_{5 \mu } &=& \xi_{5, \mu}+a^2\xi_{\mu , 5} \cr
\delta h_{55} &=& 2 \xi_5'. \cr
\da \phi &=& -\Phi_0' \xi_5 \nn
\ea
The offdiagonal $h_{5 \mu}$ components can be transformed out with
\be
 \xi_{\mu}=-\int dy \frac{1}{a^2(y)}h_{5\mu}, \; ~ \xi_{5}=0 .
\label{h5mu0}
\ee 
The branes are not moved  in the $5^{th}$ dimension.

Without loss of generality the Einstein equation can
be used without offdiagonal linear perturbation. 
The metric is
\be
ds^2 = a^2(y) (\eta_{\mu\nu}+h_{\mu\nu}(x,y)) dx^{\mu} dx^{\nu}
    -(1+2G(x,y)) dz^2.  \label{metric1}    
\ee
(\ref{h5mu0}) was our first gauge choice and there is a restricted
gauge degree of freedom to keep $h_{5\mu}=0$ .
$\xi_5(x,y)$ is constrained by the orbifold conditions 
(\ref{xi5orb}) and
\be
\xi_{\mu}(y,x)=-\int_0^y dy_1 \frac{1}{a^2(y_1)} \xi_{5, \mu}(x,y_1)+ 
\epsilon_{\mu}(x) .
\label{h5munull}
\ee
In what follows we work on the interval $(0,r_c)$ and 
get the fields on the orbifold using the $Z_2$ symmetry.

The linearized equations are the following,
'55', '$5\mu$' and  '$\mu \nu$'  respectively.
\begin{eqnarray}
\frac{\Box G}{a^2} +4 \frac{a'}{a} G' \! \! \!
  && + 2 \left (4\frac{a'^2}{a^2}+ \left ( \frac{a'}{a}  \right)' \right ) G 
-\frac{1}{2a^2}
\left ( a^2 {h}' \right )' =   \\
& &   \ka \left [ \frac{2}{3} \left ( \Phi_0''+4 \frac{a'}{a} \Phi_0' \right ) \phi +
  2\Phi_0' \phi' 
  + \frac{2}{3} \sum_i \left(\frac{\partial \lambda_i}{\partial\Phi} \phi 
  +\lambda_i G  \right ) \delta(y-y_i) \right ]  .\nn
\label{55}
\ea
\be
3  \frac{a'}{a} G_{,\mu} +
 \frac{1}{2 }\left ( h^{\alpha}_{\mu, \alpha}-h_{,\mu} \right)'  
   = \ka  \Phi_0'(y) \phi_{,\mu}(x,y) 
\label{eq5mu}
\ee
\ba
-G_{,\mu \nu} &-& \left  [ a'a G' +2\left (4a'^2+a^2 \left ( \frac{a'}{a}
    \right)' \right ) G  \right ] \eta_{\mu \nu} 
 +\frac{1}{2a^2} \left ( a^4 h_{\mu \nu}' \right )' +
\half \ \frac{a'}{a} h'  a^2 \eta_{\mu \nu} +  \cr
&& +\frac{1}{2} \left (2 h^{\lambda}_{(\mu,\nu)\lambda} 
- h^{\;\;\;\;,\lambda}_{\mu\nu,\lambda} - h_{,\mu\nu} \right ) = 
\label{eqmunu} \\
 &&  - \frac{2\ka }{3}  a^2 \eta_{\mu \nu} 
 \left ( \Phi_0''+4 \frac{a'}{a} \Phi_0' \right ) \phi 
 +\frac{\ka}{3} \sum_i a^2 \eta_{\mu \nu} \left(\frac{\partial \lambda_i}{\partial\Phi} 
   \phi  +\lambda_i G  \right ) \delta(y-y_i) 
\nn 
\ea
Here $h=h^\mu_\mu$, $\Box= \partial^\mu \partial_\mu$.
The  scalar equation is
\ba
&&\frac{1}{a^2} \Box \phi -\phi'' -4\frac{a'}{a}\phi'+
   \frac{\deriv^2 V}{\deriv \Phi^2}\phi = \cr
&&
\left (  \frac{1}{2}h'  -G' \right ) \Phi_0' - 
2G \frac{\deriv V}{\deriv \Phi}- 
\sum_i \left(   \frac{\deriv^2 \lambda_i}{\deriv \Phi^2}\phi+ 
 G \frac{\deriv \lambda_i}{\deriv \Phi} \right )   \delta(y-y_i) .
\ea
The singular terms at the branes define boundary conditions.
The complete $5\mu$ equation is valid also at the branes.
The $55$ boundary condition is the trace of
$\mu \nu$ boundary condition calculated from (\ref{eqmunu}).
It is equivalent to the Israel jump equation 
calculated with the extrinsic curvature .
From $\mu \nu$ (\ref{eqmunu}) we get
\be
h'_{\mu \nu}- 2 \frac{a'}{a} G(x,y)\eta_{\mu \nu}+ 
\frac{2 \ka}{3} \Phi_0' \phi \eta_{\mu \nu} =0 
\hbox{ at } y=y_i .
\label{bcmunu}
\ee
Already the leading order boundary conditions and $Z_2$
symmetry were used.
The scalar equation gives
\be
\pm 2 \phi'(x,y_i) =  \frac{\deriv^2 \lambda_i}{\deriv \Phi^2}\phi(x,y_i)+ 
\frac{\deriv \lambda_i}{\deriv \Phi} G(x,y_i)  .
\label{bcsc}
\ee
The equations are not idependent. The Bianchi identities ensure
that from the 15 Einstein equations are only 10 independent
and  the bulk scalar equation can be derived from the Einstein
equations with differentiation and taking linear combinations.
The scalar boundary condition presents a meaningful constraint.
The equation are solved for special BPS branes 
fulfilling (\ref{bpscond}).

\subsection{Lorentz decomposition of $h_{\mu \nu}$ }

To solve the equations we decompose $h_{\mu \nu}$ in a
Lorentz covariant way.  It contains a TT
(transverse traceless)   symmetric tensor  (5 components), one divergencefree
vector (3) and   two scalars (1+1) giving 10 components.
\be
h_{\mu \nu} =  \tilde{h}_{\mu \nu }(x,y)+ 
   A_{\mu , \nu}(x,y) + A_{\nu , \mu}(x,y)
+\partial_{\mu} \partial_{\nu} b(x,y)-F(x,y)\eta_{\mu \nu}
\ee
Here
\ba
&&\tilde{h}^{\nu}_{\, \mu, \nu }=0,~~ 
\tilde{h}^\mu_{\mu}=0,  \\
&&A^\mu_{, \mu} =0 .
\ea
The components can be uniquely defined if the
perturbations are assumed to be Lorentz covariant
as the background solution is Lorentz invariant
and fulfill
\be
\Box A_\mu \neq 0, ~~ \Box b \neq 0,
\ee
to have  nonvanishing trace or divergence in $h_{\mu \nu}$.
They are defined by
\ba
\Box^2 b &=& \frac{4}{3} \left ( \partial_\alpha \partial_\beta 
  - \Box \frac{1}{4}  \eta_{\alpha \beta} \right ) h^{\alpha \beta} ,
\label{bcomp} \\
F & =& \frac{1}{4} \left ( -h+ \Box b \right ),  \\
\Box A_\mu &= & \left ( h^{\lambda }_{\; \mu , \lambda} + F_{, \mu} 
- \Box b_{, \mu} \right ).
\label{Acomp}
\ea
This sequential definition of components ensures that
$F, \Box b$ describe the complete trace and $F,b , \Box A_\mu$
describe the whole divergence of $h_{\mu \nu}$.\footnote{ 
A subtlety of the decomposition that
$h_{\mu \nu}$ only defines $\Box^2 b$ while in the definition of
$F, A_\mu$ purely $\Box b$ appears.}
The TT part  is defined as the vector and scalar parts subtracted from 
$h_{\mu \nu}$.
$\th_{\mu \nu}$ can still contain scalar and vector TT combinations
but those will be proved to be gauge degrees of freedom as
in four dimensional gravity.

Gauge transformations of the components can be defined  as
\ba
\delta \phi &=& -\Phi_0' \xi_5(x,y),  \\
\delta G &=& - \xi_5(x,y)' ,  \label{Gg1} \\
\delta F &=& 2 \frac{a'}{a} \xi_5(x,y)  .
\ea
The last two transformation can be used to
fix $\xi_5(x,y)$ without moving the branes and
setting the $F(x,y)-G(x,y)$ difference only $x$ dependent.
Make a gauge transformation with
\ba
a^2 \xi_5(x,y) &=&
-\int_0^y a^2(y_1)(F(x,y_1)-G(x,y_1)) dy_1 
+  \chi(y) \int_0^{r_c}a^2(F-G) dy_1,  \cr
\chi(y) &=& \frac{\int_0^y a^2(y_1)dy_1 }{\int_0^{r_c} a^2(y_1)dy_1}
. \nn
\ea
This $\xi_5$ fulfills the orbifold condition
and it is completely fixed.
The gauge transformation was motivated by the work of Kubyshin et al.
who have found the radion wave function with fixed flat branes
\cite{Boos}.
In the new gauge
\be
F^n(x,y)-G^n(x,y)=f_\Da(x)=\frac{1}{\int_0^{r_c} a^2(y_1)dy_1}
 \int_0^{r_c}a^2(F^{}-G^{}) dy_1
\ee
We work in this gauge and drop the $n$ index in what follows. 
The remaining gauge degree of freedom is
\be
\xi_\mu=\ep_\mu(x),
\ee
which can be only used to set the two helicity state of the
massless four dimensional graviton. The gauge transformation
of the other components  is
\ba
\Box \da b &=& 2 \ep^\mu_{\, , \mu} ~\hbox{ , if } 
\Box \ep^\mu_{\, , \mu} \neq 0 \\
\da A_\mu &=& \ep_\mu  ~~ \hbox{ , if } 
\Box \ep_{ \mu} \neq 0, ~ \ep^\mu_{\, , \mu}=0
\ea

Untill this point the equations of motion were not used
and the gauge could be partially fixed.
In what follows we use the Einstein equations
and show that $A_\mu(x,y), b(x,y)$ can be gauged away
and $\th_{\mu \nu}$ contains only the graviton and its
KK tower. The single $F(x,y)$ field will describe
the massless radion and the scalar field together.

\subsection{Solving the Einstein equations}

The equations can be written up with the general decomposition
of $h_{\mu \nu}$. The $5 \mu$  equation (\ref{eq5mu}) gives
\be
\half \Box A_\mu' = \ka \Phi_0'(y) \phi_{, \mu}(x,y)
-\frac{3}{2}  F'_{, \mu}   - 3 \frac{a'}{a} G_{, \mu}.
\label{eq5muc}
\ee
Different Lorentz spin perturbations decouple and we get
\be
A_\mu'(x,y)=0 ,
\ee
as it was defined as $\Box A_\mu \neq 0$.
The solution is
\be
A_\mu(x,y)= f_\mu(x), ~~ \ f^\mu_{\, , \mu} \neq 0, ~ \Box f_\mu(x) \neq 0.
\ee
It can be gauged to nothing with
\be
\ep_\mu(x)= -f_\mu(x) .
\ee
The RHS of (\ref{eq5muc}) gives setting one y-dependent integration
constant zero 
\be
\frac{2 \ka}{3} a^2 \Phi_0' \phi = (a^2 G)' .
\label{phig}
\ee

In the $\mu \nu$ Einstein equation (\ref{eqmunu}) as the
result of assumed Lorentz covariance the $\pa_\mu \pa_\nu$
derivatives terms decouple from the TT and $\eta_{\mu \nu}$ part
and give the equation
\be
\frac{1}{a^2} (F-G)_{, \mu \nu} + \frac{1}{2 a^2} 
\left ( a^4 b'_{, \mu \nu} \right )' =0 .
\ee
The general solution sending two integration constant in
the infinite $x$ to zero is
\be
b(x,y)= b_1(x) + b_2(x)\int \frac{1}{a^4} + 2 f_\Da(x) 
\int\frac{1}{a^4}\int {a^2} \hbox{ , if } \Box f_\Da(x) \neq 0.
\ee
The trace of the $\mu \nu$ boundary condition (\ref{bcmunu}) reads 
\be
\Box b' - 4 \left (F' + 2 \frac{a'}{a} G -\frac{2 \ka}{3} a^2 \Phi_0' \phi
\right ) =0.
\ee
The (\ref{phig}) solution of $5 \mu$ is also valid at the branes giving
with $\Box b \neq 0$
\be
b'(x,y_i)=0.
\ee
At $y=0$ it sets $b_2(x)=0$ and at $r_c$ $f_\Da(x)=0$.
The remaining solution in $b(x,y)$ can be gauged to zero by
\be
\ep^\la_{, \la}(x) = - \half b_1(x).
\ee
If $\Box f_\Da(x) =0$ then 
using the TT part of the '$\mu \nu$' equation,
one gets $f_{\Da}(x)_{ , \mu \nu}=0$ and
$f_{\Da}(x)=0$.

We proved using the Lorentz covariance of the linear perturbations
that the following gauge can always be chosen.
There is a TT tensor and a single $F(x,y)$ scalar perturbation ($F=G$).
\ba
h_{\mu \nu} &=& \th_{\mu \nu}- F(x,y) \eta_{\mu \nu} \\
h_{55} &=& - 2 F(x,y) \\
\frac{2 \ka }{3} \Phi_0' \phi &=& F^{ \prime}(x,y) +
2 \frac{a'}{a} F(x,y)
\label{newtong}
\ea
This gauge is called Newton gauge by Tanaka and Montes
in \cite{Tanaka:2000er} and was used in \cite{Csaki}.
To our knowledge this gauge choice is correct in the presence
of two BPS branes.
The remaining gauge degree of freedom, $\Box \ep_\mu(x)=0$
can transform out the 3 non-physical component
of the massless graviton.

The scalar and the spin-2 TT perturbation decouple.
The TT equation is
\be
\frac{1}{2a^2} \left(a^4 \tilde{h}_{\mu \nu}'  \right ) '
 -\frac{1}{2} \Box  h^{TT}_{\mu\nu} =0,
\label{munuTT}
\ee
together with the boundary condition
\be
\tilde{h}_{\mu \nu}'(x,y_i) =0.
\ee
The boundary condition only allow the massless solution of
\be
\tilde{h}_{\mu \nu}=h^G_{\mu \nu}(x)
\ee
the massless graviton with 2 helicity states.
It is localized at the Planck brane with larger warp factor.
There is also a massive graviton  KK tower, with 5 degrees of freedom.
The KK modes are not localized.

Upon using the Newton gauge relation (\ref{newtong}) the Einstein boundary
conditions automatically fulfilled. In the bulk the 
scalar part of the Einstein and the 
Klein-Gordon equations are equvalent.
\be
\frac{1}{2a^2}\Box F - \half F''+
\left ( \frac{\Phi_0''}{\Phi_0'}- \frac{a'}{a} \right ) F'
+ 2 \left ( \left( \frac{a'}{a} \right )\frac{\Phi_0''}{\Phi_0'} 
-{\left( \frac{a'}{a} \right )'} \right )F =0
\label{eqrad}
\ee
This is accompanied by the scalar boundary condition (\ref{bcsc}).
Eliminating $\phi(x,y)$ it is
\be
(a^2F)''- \left ( 2 \frac{a'}{a} +\frac{\Phi_0''}{\Phi_0'}
 \pm  \half \frac{\deriv^2 \lambda_i}{\deriv \Phi^2}  \right ) (a^2F)'+
 2   \left ( \frac{a'}{a} \right )' (a^2F)=0.
\ee
This boundary condition contains a second derivative,
it makes generally the operator in the radion differential
equation non-hermitic  and causes the hermiticity problem
observed in \cite{Csaki}. 
The different mass eigenmodes are not orthogonal to each other
and does not serve as a basis for expanding a general solution.
The bulk equation can be used
to eliminate the second derivative but then the d'Alembertian appears 
with different eigenvalues for  different eigenfunctions.
This simplified boundary condition is
\be
 \left ( \frac{\Phi_0''}{\Phi_0'}
   \pm  \half \frac{\deriv^2 \lambda_i}{\deriv \Phi^2}(\Phi_0)  \right ) 
 (a^2F)' + \Box (F) =0  \; \; \; \hbox{  at }\; y=0,r_c .
\ee
It is generally still not hermitic, though with BPS branes the first
term vanishes  as 
$\frac{\Phi_0''}{\Phi_0'}= \half \frac{\deriv^2 W}{\deriv \Phi^2}$.
The
\be
\Box F(x,y_i)=0
\label{bcscalar}
\ee
boundary condition defines a hermitic operator for massive perturbations.
Practically there is no boundary condition for the massless solutions.

\subsection{Massless  scalar perturbations}

There are  two independent  massless modes 
as the boundary condition (\ref{bcscalar}) is
automatically fulfilled.
The radion differential equation can be written as
\be
\frac{\left( \frac{a'}{a} \right )'}{\left( \frac{a'}{a} \right )}
\left (  \frac{1}{a^2} \frac{\left( \frac{a'}{a} \right )^2}
{\left( \frac{a'}{a} \right )'} \left ( a^2 \frac{1}{\frac{a'}{a}  }F  \right )'
\right )' - \frac{\Box}{a^2} F=0
\label{radion0}
\ee
The wavefunction of the  massless radion is
\be
F_R(x,y)=F_R(y) r(x)=\ep(y)  \frac{a'}{a } \frac{1}{a^2(y)} r(x) 
  \quad \Box r(x)=0,
\label{radionwavef}
\ee
where $\ep(y)$ is the sign function and  it makes the radion wavefunction
even under $Z_2$. 
The radion appears in the scalar field perturbation as
$\phi_R(x,y)=-\frac{\Phi_0'}{2a^2}r(x)$.
In the limit of no scalar field (\ref{radionwavef})
gives back the result of Charmousis et al. \cite{cgr}
$ F_R= -k e^{2 ky} r(x)$.
The radion describes the relative free motion of the branes.
With a  gauge transformation of 
\be
\xi_5(x,y)= -\frac{1}{2a^2}  r(x),
\ee
the radion solution is completely transformed to 
a scalar TT mode and the relative motion of the branes. 
Keeping $h_{5 \mu}=0$ with (\ref{h5munull}) after
this gauge transformation the bulk metric is
\ba
h_{\mu \nu} &=& \th_{\mu \nu}+r_{, \mu \nu}(x)\int_0^y \frac{1}{a^4}
- F_s(x,y) \eta_{\mu \nu} \\
h_{55} &=& - 2 F_s(x,y) \\
\frac{2 \ka }{3} \Phi_0' \phi_s &=& F_s^{ \prime}(x,y) +
2 \frac{a'}{a} F_s(x,y).
\ea
$F_s, \phi_s$ here only describes the scalar field perturbations.
With BPS branes this gauge transformation is well defined
and the relative motion of the branes is
\be
\Da r_c=\half \left ( \frac{1}{a^2(0)}-\frac{1}{a^2(r_c)} \right ) r(x).
\ee
This single radion solution was analyzed in \cite{Brax:2002kv}
without the dynamics of the scalar field.

The scalar field has a zero mode and a complete KK tower.
The even massless scalar solution of (\ref{radion0}) 
with flat branes is
\be
F_D(x,y)= F_D(y) f_1(x)=\frac{a'}{a } \frac{1}{a^2(y)} f_1(x) 
\int_0^y \deriv y_1 a^2 \frac{\left( \frac{a'}{a }\right )' }{\left( \frac{a'}{a}
  \right )^2}\quad , \quad \Box f_1(x)=0 . 
\label{dilaton}
\ee
This is the dilatonic zero mode of the scalar field,  it is the 
Goldstone boson of the spontaneously
broken dilatational invariance of the action.
It appeared as adjusting  a meaningless constant in the limit 
of the Randall-Sundrum scenario with decoupled scalar field.
Brax et al. did not discuss  this 
second zero mode in  \cite{Brax:2002kv}.
They were aware of two scalar zero modes \cite{bdrb} and recently
studied the cosmological evolution of them \cite{Brax:2002nt}.

The extra dilatonic zero mode given in (\ref{dilaton}) 
vanishes on the Planck brane
and its  wavefunction is  the radion wavefunction multiplied
with an extra y-function. 
The two solutions are not orthogonal to each other,
they can be orthogonalized  with Gram-Schmidt method,
once  a scalar product is defined  on the  interval $(0,r_c)$.
Two solutions can be distinguished also on physical grounds.
The dilatonic zero mode (\ref{dilaton})
also describes a relative motion of the branes through 
changing the proper distance 
$\int_0^{r_c}\sqrt{|g_{55}|} \deriv y= r_c+\int_0^{r_c}F(x,y) \deriv y$.
A new dilaton wavefunction can be formed with linear combination
that keeps the proper distance $r_c$ unchanged.
\be
F_2(y)=F_D(y)-\frac{\int_0^{r_c}F_D(y)}{\int_0^{r_c}F_R(y)} F_R(y)
\ee
Now the two massless modes have different physical origin
as one moves the branes the other not.
The radion orthogonal to  $F_2(y)$ 
with respect to a specific scalar product can be defined as
$\tF_R(y)= F_R(y)-F_2(y) (F_R,F_2)/(F_2,F_2)$.

The dilatonic zero mode remarkably simplify with exponential superpotential.
Performing the integral in the solution (\ref{dilaton})
we get
\be
F_D(x,y)=-f_1(x) \frac{1}{1+\frac{2 \ka}{3 \a^2}} \left( \frac{1}{k}
\frac{a'}{a} \frac{1}{a^2}+1 \right )
\simeq  f_1(x) (F_R(y)-F_R(0)).
\ee
Subtracting the radion part we get 
$F_D(x,y)= f_1(x)/({1+\frac{2 \ka}{3 \a^2}})$ 
which implies via (\ref{newtong}) the
dilaton scalar field solution
\be
\phi_D(x,y)=-\frac{1}{\a} \frac{1}{1+\frac{2 \ka}{3 \a^2}}f_1(x).
\label{dilaton2}
\ee
In the limit $\a \rightarrow 0$ it is the dilaton of (\ref{RSdilaton}).
This solution can be found directly as the coefficient of 
the term $F$ in the radion  equation (\ref{eqrad}) 
vanishes for $W \sim e^{\a \Phi}$,
\be
 \left ( \left( \frac{a'}{a} \right )\frac{\Phi_0''}{\Phi_0'} 
-{\left( \frac{a'}{a} \right )'} \right )=\frac{\ka}{12} W^2
\left( \frac{W'}{W} \right )' =0, 
\hbox{ here } W'= \frac{\pa W}{\pa \Phi}.
\ee
Thus the equation (\ref{eqrad}) has a $y$ independent $F_D(x,y) \simeq f_1(x)$
dilaton solution of  (\ref{dilaton2}).

The dilatonic zero mode, called $\omega_2$ 
was found by Bozza et al. \cite{Bozza:2001xt}
with one brane in the case of  dilatonic superpotential.
This was the only propagating massless mode as the radion
$\om_3$ is not normalizable and not physical with one brane.
$\om_2$ was identified as the dilaton in the limit of the
RS1 scenario when the scalar field decouples from gravity.
The Einstein equations gave constraints between the massive modes
and there were two independent scalar KK tower.
One is the scalar field KK tower. The other completes
the two component transverse traceless tensor $\bar{h}_{ij}$
with two additional 3-vector mode to a 5 componenent massive 
4-dimensional KK graviton.
Bozza et al. investigated the scalar perturbation of a  Lorentz 
invariant setup based on 3 dimensional spatial covariance.
At the end  4 dimensional covariant equations described the
perturbations. It would be enlightening to perform their
analysis in 4 dimensional language and understand the physics
of perturbations. Indeed following their method 
with two branes  we also got two massless scalar modes.
This work is in progress.

\subsection{Massive KK tower}

There is a KK tower of the massive scalar field perturbations
with the well defined hermitic boundary conditions $F(x,y_i)=0$.
In a scenario with small backreaction like in \cite{Csaki},
the KK spectrum can be solved with good approximation
using the  exponential warp factor $a(y)= e^{-ky}$.
The mass eigenvalues determined by the zeros of the Bessel functions
and are at the order of the weak scale ($ke^{-kr_0}$).

The radion equation (\ref{eqrad}) for the case of
exponential superpotential (\ref{dilatonpot})
can be analytically solved by Bessel functions. 
As a standard technique we change to conform flat background
by $dy= a(z) dz$ and  rescale the field F by
$F(x,z)=e^{ipx} \Phi_0' a^{-\frac{3}{2}} \tF(z)$
also changing to mass eigenvectors with
$\Box F_i=-m_i^2 F_i$.
The radion equation in this Schr\"odinger basis is
\be
\tF''-\left( \frac{9}{4} \left ( \frac{a'}{a} \right )^2
-\frac{5}{2} \left ( \frac{a'}{a} \right )'+
\frac{a'}{a} \frac{\Phi_0''}{\Phi_0'}+
2\left( \frac{\Phi_0''}{\Phi_0'}\right )^2-\frac{\Phi_0'''}{\Phi_0''}
 \right ) \tF+ m^2 \tF =0 .
\ee
This equation for each eigenmode has a mass parameter and 
two integration constants. One integration constant is
fixed by the normalization, the other from the boundary
condition at the Planck brane ($y=0$) and the mass is
determined from the TeV brane ($r_c$) boundary condition.
With the exponential superpotential the equation is
\be
\tF''+\left (m^2 - 
\frac{ \nu^2-\frac{1}{4}}{\zeta^2} \right )\tF =0,
\ee
where
\ba
\zeta &=& \frac{1}{k(1-\frac{3 \a^2}{\ka}) }+ z= \frac{1}{\tk}+z, \\
\nu &=& -\half -\frac{3}{2} \frac{1}{\frac{3 \a^2}{\ka} -1}.
\ea
The range of $\nu$ is $1\leq  \nu < \infty$, $\nu=1$ corresponding
to the RS scenario.
The solution fulfilling the $y=z=0$ boundary condition is
given by two  type of Bessel functions
\be
\tF = \sqrt{\zeta} \left( Y_{\nu}\left (\frac{m}{\tk}\right ) 
J_{\nu}\left (m \zeta \right )
- J_{\nu} \left (\frac{m}{\tk} \right ) 
Y_{\nu} \left (m \zeta \right ) \right ) .
\ee
The mass eigenvalues are determined at $y=r_c$ from the equation
\be
J_{\nu} \left (\frac{m_i}{\tk} \right )
\frac{Y_{\nu} (m_i \zeta_c)}
{Y_{\nu}\left (\frac{m_i}{\tk} \right )}
-J_{\nu} \left ( m_i \zeta_c \right )=0 ,
\ee
with
\be
\zeta_c=\frac{1}{\tk}(1-\frac{3 \a^2}{\ka}k r_c)^\frac{\ka}{3 \a^2}
= \frac{a(r_c)}{\tk}.
\ee
There is always an $m_i=0$ eigenvalue, a massless solution.
When the warp factor generates a large hierarchy 
between the two branes $a(r_c) \ll a(0)=1$ then the lowest 
mass eigenvalues are practically determined from the second term,
the zeros of the Bessel function $J_{\nu} \left ( m_i \zeta_c \right )$.
The lowest KK masses are  few time  the warped mass scale
$\tk a^{-1}(r_c)$, the TeV scale.

$\tF$ uniquely defines $\tp$ via (\ref{newtong}) and
\be
\tp \simeq \left ( \frac{6 \a^2}{ \ka m} \frac{1}{\sqrt{\zeta}} J_\nu(m \zeta)
-\left (1+\frac{3 \a^2}{\ka}\right ) J_{1+\nu}(m \zeta) \right ) 
Y_\nu(\frac{m}{k}) - (Y \leftrightarrow J)
\ee
In the limit $\a=0, \nu=1$ we get back the masive spectrum given in
(\ref{RSphi}).

\section{Conclusion}

In this paper we have analyzed the coupled scalar gravitational system in 
5 dimensions. Around an exact background solution we 
carefully fixed the gauge and solved
the 5 dimensional linear equations of motion on the orbifold.
We have found that with the special BPS brane potentials there are
two scalar zero modes and one KK tower. The two scalar zero modes
are the radion and the dilaton which can be  identified in
the smooth limit of the RS1 scenario with a free bulk scalar field.
Both modes couple to the matter on the TeV brane via the
induced metric. 
The two massless scalar modes generally 
lead to unacceptable phenomenology,
the radius is not stabilized. 
Cosmological evolution of two scalar moduli in this model were
discussed recently \cite{Brax:2002nt}.
However the BPS brane scenario
may help us to  find the correct linear spectrum in
the Goldberger-Wise stabilization mechanism.
The gauge fixing discussed in this paper and used in \cite{Csaki}
\cite{Tanaka:2000er} leads to non-hermitic differential operator
for scalar perturbations for general brane potentials.
The GW bulk scalar with detuned potential gives mass not just
to the radion but also to the dilaton mode. It would be very interesting
to follow the fate of these two modes and relating them to the
first scalar KK mode.
Effective theory and AdS/CFT based calculation so far
could only capture the radion mode \cite{Goldberger:1999un},
\cite{kanti}, \cite{rattazzi}.
The possible resolution of the hermiticity problem is that
a mode is droped during solving the Einstein equations which has
the  the detuning of the  brane potentials as a source.
This work is in progress.

\vspace{0.5cm}
{\large \bf Acknowledgement}
\vspace{0.3cm}

The author is grateful to Zolt\'an Kunszt, Csaba Cs\'aki and
Martin Puchwein for useful conversation. This
work was supported in part by SNF and OTKA 029803.

\end{document}